\documentclass[aps,prl,twocolumn,unsortedaddress,superscriptaddress,nobibnotes]{revtex4}
\usepackage{color,graphicx}
\usepackage{array,amsmath,amssymb,pifont,mathrsfs,latexsym}
\usepackage{lipsum}
\usepackage{epstopdf}

\begin{document}

\title{Effective Nucleon Masses from Heavy Ion Collisions}
\author{D.D.S. Coupland}
\affiliation{National Superconducting Cyclotron Laboratory, Michigan State University, East Lansing, MI 48864, USA}
\affiliation{Department of Physics and Astronomy, Michigan State University, East Lansing, MI 48864, USA}
\author{M. Youngs}
\affiliation{National Superconducting Cyclotron Laboratory, Michigan State University, East Lansing, MI 48864, USA}
\affiliation{Department of Physics and Astronomy, Michigan State University, East Lansing, MI 48864, USA}
\author{W.G. Lynch}
\affiliation{National Superconducting Cyclotron Laboratory, Michigan State University, East Lansing, MI 48864, USA}
\affiliation{Department of Physics and Astronomy, Michigan State University, East Lansing, MI 48864, USA}
\affiliation{Joint Institute of Nuclear Astrophysics, Michigan State University, East Lansing, MI 48864, USA}
\author{M.B. Tsang}
\affiliation{National Superconducting Cyclotron Laboratory, Michigan State University, East Lansing, MI 48864, USA}
\affiliation{Department of Physics and Astronomy, Michigan State University, East Lansing, MI 48864, USA}
\affiliation{Joint Institute of Nuclear Astrophysics, Michigan State University, East Lansing, MI 48864, USA}
\author{Z. Chaj\c{e}cki}
\affiliation{National Superconducting Cyclotron Laboratory, Michigan State University, East Lansing, MI 48864, USA}
\author{Y. X. Zhang}
\affiliation{China Institute of Atomic Energy, P.O. Box 275 (10), Beijing 102413, P.R. China}
\author{M.A. Famiano}
\affiliation{Department of Physics, Western Michigan University, Kalamazoo, MI 49008, USA}
\author{T.K. Ghosh}
\affiliation{Variable Energy Cyclotron Centre, 1/AF Bidhannagar, Kolkata 700064, India}
\author{B. Giacherio}
\affiliation{Department of Physics, Western Michigan University, Kalamazoo, MI 49008, USA}
\author{M. A. Kilburn}
\affiliation{National Superconducting Cyclotron Laboratory, Michigan State University, East Lansing, MI 48864, USA}
\affiliation{Department of Physics and Astronomy, Michigan State University, East Lansing, MI 48864, USA}
\author{Jenny Lee}
\affiliation{National Superconducting Cyclotron Laboratory, Michigan State University, East Lansing, MI 48864, USA}
\affiliation{Department of Physics and Astronomy, Michigan State University, East Lansing, MI 48864, USA}
\author{F. Lu}
\affiliation{Joint Institute of Nuclear Astrophysics, Michigan State University, East Lansing, MI 48864, USA}
\affiliation{Shanghai Institute of Applied Physics, Chinese Academy of Sciences, Shanghai 201800, China}
\author{P. Russotto}
\affiliation{INFN, Sezione di Catania, I-95123 Catania, Italy}
\author{A. Sanetullaev}
\affiliation{National Superconducting Cyclotron Laboratory, Michigan State University, East Lansing, MI 48864, USA}
\affiliation{Department of Physics and Astronomy, Michigan State University, East Lansing, MI 48864, USA}
\author{R. H. Showalter}
\affiliation{National Superconducting Cyclotron Laboratory, Michigan State University, East Lansing, MI 48864, USA}
\affiliation{Department of Physics and Astronomy, Michigan State University, East Lansing, MI 48864, USA}
\author{G. Verde}
\affiliation{INFN, Sezione di Catania, I-95123 Catania, Italy}
\author{J. Winkelbauer}
\affiliation{National Superconducting Cyclotron Laboratory, Michigan State University, East Lansing, MI 48864, USA}
\affiliation{Department of Physics and Astronomy, Michigan State University, East Lansing, MI 48864, USA}

\date{\today}

\begin{abstract}
We probe the momentum dependence of the isovector mean-field potential by comparing the energy  spectra of neutrons and protons emitted in $^{112}$Sn+$^{112}$Sn and  $^{124}$Sn+$^{124}$Sn collisions at incident energies of E/A=50 and 120 MeV. We achieve experimental precision that discriminates between different momentum dependencies for the symmetry mean-field potential.
Comparisons of the experimental results to Improved Quantum Molecular Dynamics model calculations with Skyrme Interactions indicate small differences between the neutron and proton effective masses.
\end{abstract}

\maketitle

The properties of neutron-rich matter and its Equation of State (EoS) have important connections to the structure and stability of exotic neutron-rich nuclei, core-collapse supernovae and neutron stars~\cite{steiner05,janka12,bethe90,dan03s,dan09}.
To delineate these connections, one needs to know the EoS as a function of temperature and of the densities of neutrons $\rho_{n}$ and protons $\rho_{p}$ for values of $\rho =\rho_{n}+\rho_{p}$ extending well beyond the saturation value of $\rho_{0}\approx 2.7\times10^{14} g/cm^{3}$. Measurements of collective flow and kaon production in heavy ion collisions (HIC) have provided constraints on the EoS of \emph{symmetric} matter ($\rho_{n}=\rho_{p}$) at $\rho/\rho_{0}=2-4.5$ ~\cite{schmah05,dan02s,lynch09}. Disparate constraints on the EoS for \emph{neutron-rich} matter ($\rho_{n} \gg \rho_{p}$) and $\rho \approx 2.5 \rho_{0}$ have been obtained from neutron star radii extracted from x-ray astronomical observations ~\cite{lattimer01,lattimer12,steiner10,0004-637X-772-1-7} and from nucleus-nucleus collisions \cite{xiao09,russotto11}. The difference between the EoS for neutron matter and that for symmetric matter defines the symmetry energy. Its uncertain density dependence reflects contributions from three neutron and other higher order interactions ~\cite{PhysRevC.82.014314,PhysRevC.85.032801}, motivating the search for constraints using measurements of nuclear structure~\cite{typel01,chen10,steiner05,horowitz01,trippa08,klimkiewicz07,carbone10,abrahamyan12},
and reactions~\cite{famiano06,tsang04,liu07,souza08,tsang01,baoan01,tsang09,tsang12,reisdorf07,ferini05,xiao09,feng10,russotto11,baoan08,lafevre05,mcintosh13,Kohley13}.

 Such constraints require an improved understanding of how the nucleonic mean-field potential depends on both density and on the momenta of  nucleons~\cite{li04,liu02}. Mean-field potentials acquire momentum dependencies due to non-localities and momentum dependencies of the nucleon-nucleon interaction, and from exchange and higher-order terms ~\cite{brueckner55,mahaux85,li04,liu02,dobaczewski99,bethe90,farine01,zuo99,zuo02,hofmann01,greco03}. Over a restricted momentum range, this dependence can be approximated by replacing bare nucleon masses by effective masses~\cite{brueckner55}. Deeply bound nuclear states~\cite{mahaux85} and nucleon elastic scattering optical potentials~\cite{PhysRevC.63.024607} in symmetric matter ($\rho_{n} = \rho_{p}\approx0.5 \rho_{0}$) require nucleon effective masses of $m^{*} \approx 0.7\cdot m$. Similar conclusions were drawn from relativistic nucleus-nucleus collisions~\cite{dan00}.  However, questions about momentum dependencies remain~\cite{Hong:2013yva}, especially concerning isovector contributions contained in the symmetry mean-field potentials that can make the neutron and proton effective masses different~\cite{li04,liu02}.

Differences between the neutron and proton effective masses strongly influence the symmetry energy, the thermal properties of neutron-rich nuclei and neutron stars~\cite{Pons:1998mm},
and the magnitude of shell effects in nuclei far from stability~\cite{dobaczewski99,bethe90,farine01}.
Calculations using Landau-Fermi liquid
theory~\cite{sjoberg76} and the non-relativistic Brueckner-Hartree-Fock~\cite{zuo99,zuo02}
approach have predicted that $m^*_n > m^*_p$
in neutron rich matter while relativistic mean field (RMF) and other calculations using relativistic Dirac-Brueckner calculations~\cite{hofmann01,liu02,greco03} predict that $m^*_n < m^*_p$.  Analyses of nucleon-nucleus elastic scattering somewhat prefers $m^*_n > m^*_p$ ~\cite{PhysRevC.63.024607}, but the uncertainties are large. Consequently, the sign and magnitude of the effective mass splitting are not well
constrained - especially for $\rho\neq\rho_{0}$.

 In central collisions of neutron-rich nuclei, matter in the participant region formed by the overlap of projectile and target becomes compressed and then expands reflecting properties of the EoS at $\rho\neq\rho_{0}$. Transport model calculations predict that fast neutrons from the compressed participant region will experience a more repulsive potential and a higher acceleration for $m^*_n < m^*_p$, than do fast protons at the same momentum, resulting in an enhanced ratio of neutron over proton (n/p) spectra at high energies ~\cite{Rizzo:2005mk,DiToro:2010ku,Zhang2014186,PhysRevC.88.061601}.  In contrast, calculations for $m^*_n > m^*_p$ predict that the effective masses enhance the acceleration of protons relative to neutrons resulting in a lower n/p spectral ratio ~\cite{Rizzo:2005mk,DiToro:2010ku,Zhang2014186,PhysRevC.88.061601}. In this letter, we test these predictions by presenting the first measurements of n/p spectral ratios with sufficient precision to distinguish between the theoretically predicted n/p ratios for different effective mass splittings ~\cite{Zhang2014186}.

We investigate these issues by
measuring transverse neutron and proton spectra from central $^{124}$Sn+$^{124}$Sn and $^{112}$Sn+$^{112}$Sn collisions
at incident energies of 50 MeV/u and 120 MeV/u.  The $^{124}$Sn and $^{112}$Sn  beams were accelerated to 50 and 120 MeV/u  by the Coupled Cyclotron Facility  and impinged upon $5 mg/cm^{2}$ $^{112}$Sn and $^{124}$Sn foil targets located within a thin-walled aluminium chamber in the S2 vault of the National Superconducting Cyclotron
Laboratory.

Hydrogen and helium isotopes were detected and isotopically resolved
by six $\Delta$E-E charged-particle telescopes from the Large Area Silicon Strip Array (LASSA)~\cite{davin09} placed 20 cm from the target.
Each LASSA telescope consisted
of a 500 $\mu$m double-sided $\Delta$E silicon strip detector (DSSD) backed by an E detector consisting of four 6 cm thick CsI(Tl) crystals
arranged in quadrants. These six telescopes
spanned the polar angle range $23^{\circ} < \theta_{lab} < 57^{\circ}$ with a 0.9$^\circ$ angular resolution.

Neutrons were detected by the two walls of the MSU Large Area Neutron Array (LANA)~\cite{zecher97}, placed across the beam axis from LASSA at 5 m and 6 m from the
reaction target. The LANA spanned polar angles of $15^{\circ} < \theta_{lab} < 58^{\circ}$ with an angular resolution of 0.8-0.9$^\circ$ \cite{couplandprivate}. Neutrons were distinguished from $\gamma$-rays by pulse shape discrimination and from charged particles by use of a charged-particle veto array of BC-408 plastic scintillator detectors placed between the target and neutron walls. Neutron kinetic energies were
determined by time of flight, measured with 1 ns resolution, using a start time supplied by an array of thin NE-110 plastic scintillators located 10 cm downstream from the target. The background from secondary scattering of neutrons from the floor and other materials was determined via shadow bar measurements.

\begin{figure*}
\centering
\includegraphics[width=.75\textwidth]{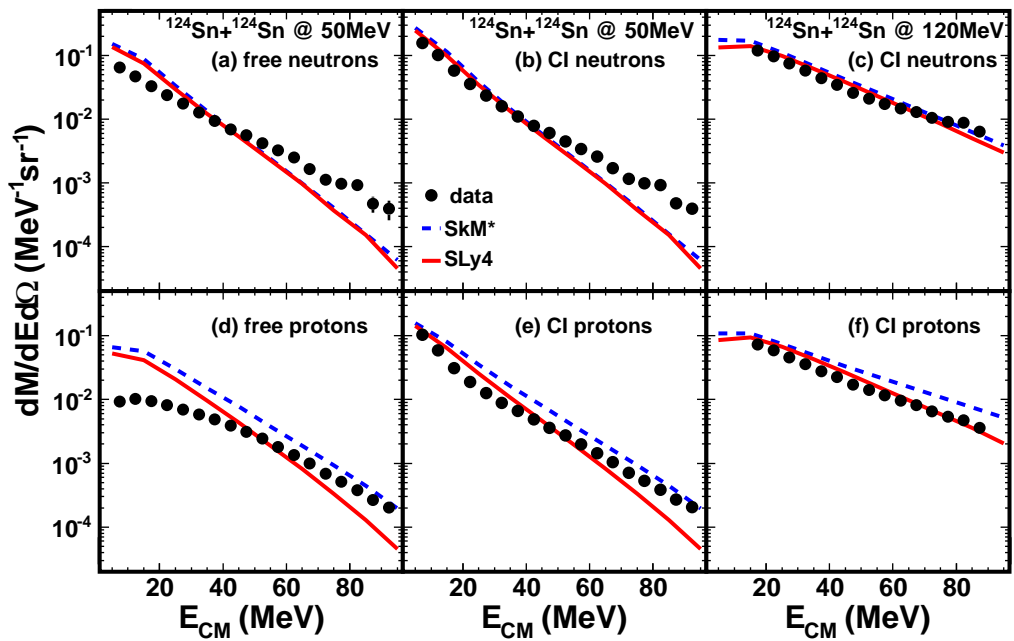}
\caption{Free neutron (a) and  proton (d), and coalescence invariant neutron (b) and proton (e) spectra from $^{124}$Sn+$^{124}$Sn reaction at 50 MeV/u and coalescence invariant neutron (c) and proton (f) spectra from $^{124}$Sn+$^{124}$Sn reaction at 120 MeV/u (black data points) compared to ImQMD-Sky calculations.  }

\label{fig:12450}
\end{figure*}

 In central collisions at this energy, most of the particles  are direct participants in the reaction, forming a mid-rapidity region that is first compressed, then rebounds and expands, emitting nucleons and light clusters in the process. Nucleons and light clusters from the participant region were detected by the MSU Miniball~\cite{desouza90}, an array of phoswich detectors that covered 70\% of the lab frame solid
angle in the chosen configuration.  The Miniball primarily measured the multiplicity and transverse energy of charged particles emitted from this participant region, which is on average a monotonically decreasing function of impact parameter~\cite{cavata90}. Gating on the highest ~6\% of the multiplicity distribution, we selected central reactions with impact parameters less than 3 fm, thereby ensuring that most $Z\leq2$ particles detected in LASSA and LANA are participants in the collision.

To further ensure that light particles detected in LASSA and LANA
are emitted from the participant region, an angular cut of $70^{\circ} < \theta_{CM} < 110^{\circ}$ is applied to select particles emitted in the transverse direction. Over this angular domain the center-of-mass energy spectra for these particles remains independent of angle~\cite{youngsprivate}. Studies show that the shapes and, consequently, the ratios (e.g. $\frac{dM_{n,124}}{dE_{CM}d\Omega_{CM}}$ $/$ $\frac{dM_{p,124}}{dE_{CM}d\Omega_{CM}}$) of the theoretical $Z\leq2$ spectra within this angular cut do not change significantly with
the impact parameter selection~\cite{Zhang:2010th}; likewise, the shapes and ratios of our experimental spectra are insensitive  to variations of the experimental impact parameter range from $ 0<b<3 fm$ to $0<b<6 fm$.

Neutron and proton center-of-mass energy spectra, $\frac{dM_{n,124}}{dE_{CM}d\Omega_{CM}}$ and $\frac{dM_{p,124}}{dE_{CM}d\Omega_{CM}}$, from the $^{124}$Sn+$^{124}$Sn reaction at 50 MeV/u are shown as the solid points in the top and bottom left panels of
Fig.~\ref{fig:12450}. These spectra are normalized to provide the differential multiplicities of neutrons and protons, respectively. The single-neutron detection efficiency was evaluated using the SCINFUL-QMD Monte Carlo code~\cite{satoh06}, which calculates the efficiency of NE-213 scintillators to an estimated accuracy of 15\%.  Known systematic and statistical uncertainties in the  spectra are smaller than the data points in this figure. As many transport models have difficulty reproducing the relative abundances of light isotopes produced as the system expands and disassembles, we calculate the coalescence invariant (CI) neutron and proton spectra by combining the free nucleons with those bound in light isotopes with $1<A<5$.  The solid points in the middle and right panels  of Figure~\ref{fig:12450} indicate the corresponding neutron and proton CI spectra at E$_{beam}$/A $=$ 50 and 120 MeV, respectively, which are constructed by adding the protons and neutrons in light clusters to the free nucleon spectra as follows
\begin{eqnarray}
  \frac{dM_{n,CI}}{dE_{CM}d\Omega_{CM}} &=& \sum_{N,Z}N\cdot\frac{dM(N,Z)}{d(E/A)_{CM}d\Omega_{CM}}\cr
  \frac{dM_{p,CI}}{dE_{CM}d\Omega_{CM}} &=& \sum_{N,Z}Z\cdot\frac{dM(N,Z)}{d(E/A)_{CM}d\Omega_{CM}}
  \end{eqnarray}
\noindent where $\frac{dM_{n,CI}}{dE_{CM}d\Omega_{CM}}$ and $\frac{dM_{p,CI}}{dE_{CM}d\Omega_{CM}}$ denote the coalescence invariant neutron and proton spectra, $\frac{dM(N,Z)}{d(E/A)_{CM}d\Omega_{CM}}$ denotes the measured differential multiplicity spectrum for fragments of charge and neutron numbers Z and N. The summation includes n,p,d,t,$^{3}He$ and $^{4}He$.

While the free proton and neutron spectra were cleanly measured up to energies of $E_{CM}/A$=100 MeV/u, the upper limits in $E_{CM}/A$ of our measured clusters  are lower, reflecting their ranges in the 6 cm thick CsI(Tl) crystals of LASSA. The small contributions of $A\geq3$ clusters that penetrate through LASSA can be neglected. However, the deuteron contributions to the coalescence invariant spectra beyond $E_{CM}/A \sim 55$~MeV were extrapolated to higher energies via the coalescence approximation, i.e.
\begin{eqnarray}
\frac{dM_{d}}{d(E/A)_{CM}d\Omega_{CM}}&=& C\cdot\frac{dM_{n}}{dEd\Omega_{CM}}\cdot\frac{dM_{p}}{dEd\Omega_{CM}}
\end{eqnarray}
\noindent where C is a normalization factor determined by matching the product on the right to the measured deuteron spectrum~\cite{Chajecki:2014vww}. The systematic uncertainty of this extrapolation to $E_{CM}/A>55$~ MeV is less than 1\% and 2.5\% at $E_{beam}/A=50$~MeV and $120$~MeV, respectively.

To illustrate the sensitivity of such data to the isospin dependent effective masses, we compare them to transport model calculations from the ImQMD-Sky quantum molecular dynamics transport model~\cite{Zhang2014186}. As discussed in ref.~\cite{Zhang2014186}, the mean-field potential is calculated using Skyrme effective interactions, which facilitates the study of the effective mass splitting. 
Here, we focus on calculations that employ the SkM* and SLy4 Skyrme potentials, which have a similar slope of the symmetry energy (L) but opposite mass splitting at saturation density as shown in Table I. At $\rho = \rho_{0}$ and $\delta=(\rho_n-\rho_p)/\rho_0=0.2$, the SkM* potential has $m^*_n > m^*_p$ with a fractional isovector mass correction $f_{I}=m_{n}/m^*_n-m_{p}/m^*_p=-0.096$, while the SLy4 potential has $m^*_n < m^*_p$ with $f_{I}=0.062$. The calculations were performed at the impact parameter of 2 fm.  Bound nuclei are identified by the proximity of nucleons to each other in position and momentum space in the final stages of the simulation.

Calculated free and CI nucleon spectra are shown in  Figure 1. The solid lines and dashed lines correspond to calculations performed with the SLy4 and SkM* mean fields, respectively. The difference in the effective masses appears to mainly influence the calculated proton spectra. At these energies, the CI data and calculations are in better agreement than the free data and calculations. This reflects the fact that these calculations underpredict the yields of bound nuclei (especially $^{4}He$), due in part to their underpredicting the binding energies of such light nuclei; thus, nucleons in $^{4}He$'s can be predicted to be emitted as free nucleons. We also find that the calculations overpredict the CI neutron and proton data at low energies and have a steeper energy dependence. The disagreement at E/A $<$ 20 MeV may be influenced by the neglect of fragments with Z$>$3, which were not measured in this experiment. The discrepancies in slope at E/A$ >40$~MeV, however, suggest inaccuracies in the treatment of the isoscalar dynamics and the symmetric matter EoS that influences both n and p spectra.
Better agreement between data and calculations is observed at $E_{beam}/A=120$~MeV.

\begin{table}
  \centering
  \begin{tabular}{|l|c|c|c|c|}
  \hline
  Skyrme & $S_{0}$(MeV) & L (MeV) & $m^{*}_{n}/ m_{n}$& $m^{*}_{p}/ m_{p}$ \\
    \hline
  SLy4 & 32 & 46 & 0.68 & 0.71 \\
    \hline
  SkM* & 30 & 46 & 0.82 & 0.76 \\
  \hline
  \end{tabular}
  \caption{Parameters of the Skyrme mean filed potentials for the calculations shown in Figs. 1 and 2. The effective mass for neutron and proton are obtained for isospin asymmetric nuclear matter with $\delta$ = 0.2.}
  \label{table1}
\end{table}

Sensitivity to the symmetry energy can be enhanced and isoscalar dynamics suppressed by dividing $\frac{dM_{n,CI}}{dE_{CM}d\Omega_{CM}}$ by  $\frac{dM_{p,CI}}{dE_{CM}d\Omega_{CM}}$ for each reaction to obtain the n/p spectral ratio $R_{n/p}(CI)$ for each reaction. One can further isolate effects due the difference between neutron and proton effective masses by constructing the coalescence invariant double ratios
\begin{equation}
\label{eq:dratio}
DR(n/p)= \frac{R(n/p,CI,124)}{R(n/p,CI,112)}
\end{equation}
\noindent as a function of $E_{CM}$. Here, $R(n/p,CI,124)$ is given by
\begin{equation}
\label{eq:ratio}
R(n/p,CI,124) = \frac{dM_{n,CI}(124)}{dE_{CM}d\Omega_{CM}}/\frac{dM_{P,CI}(124)}{dE_{CM}d\Omega_{CM}}
\end{equation}
\noindent and $R(n/p,CI,112)$ is obtained via an analogous expression involving $^{112}$Sn+$^{112}$Sn spectra. Calculations indicate that DR(n/p) is independent of the isoscalar mean-field potential, the isoscalar nucleon effective mass and the magnitudes of the isoscalar in-medium nucleon-nucleon cross sections ~\cite{youngsprivate,Zhangpriv14}. Moreover, the double ratio minimizes sensitivities to the charge distribution at breakup, to the uncertainties in neutron detector efficiencies and to the neutron and proton energy calibrations.

\begin{figure}
\centering
\includegraphics[width=0.45\textwidth]{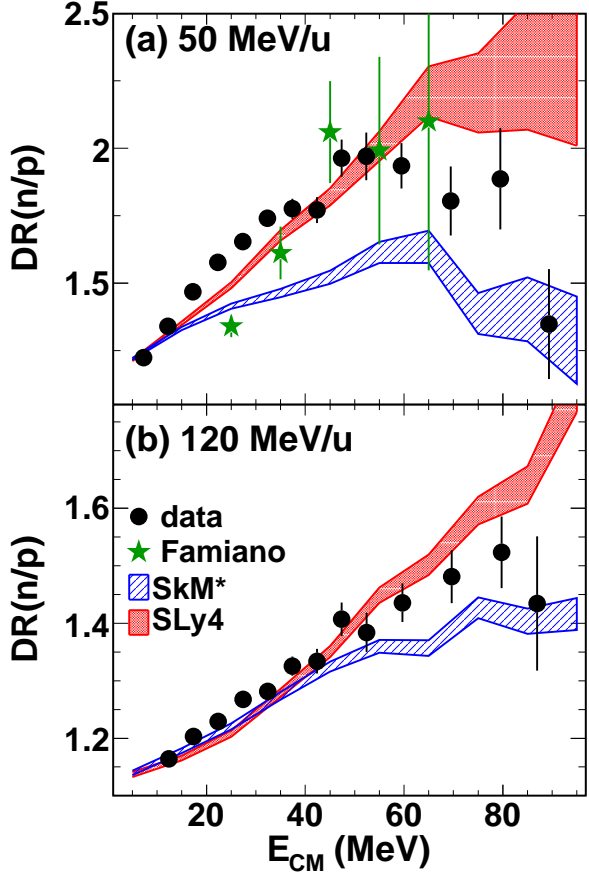}
\caption{Neutron to proton double ratio (Eq.~\ref{eq:dratio}) for $Sn+Sn$ collisions  at 50 MeV/u (a) and 120 MeV/u (b).  }
\label{fig:ciDR}
\end{figure}

Figure \ref{fig:ciDR} indicates the CI double ratios DR(n/p) for both beam energies, which can be constructed from the $\frac{dM_{n,CI}(124)}{dE_{CM}d\Omega_{CM}}$ and   $\frac{dM_{p,CI}(124)}{dE_{CM}d\Omega_{CM}}$ spectra in Figure 1 for the $^{124}$Sn+$^{124}$Sn reaction and corresponding spectra for the $^{112}$Sn+ $^{112}$Sn reaction (not shown). There are systematic uncertainties in DR(n/p) of about 10\% at $E_{beam}/A = 50$~MeV and 15\% at $E_{beam}/A = 120$~MeV (not shown in Figure \ref{fig:ciDR}) stemming from the dependence of the neutron detection efficiencies
on the charged particle and scattering background in LANA.
Previous double ratio data from Ref. ~\cite{famiano06} are indicated by the green stars in the 50 MeV/u panel.  That data set had an impact parameter cut of $b < 5$ fm, compared to $b < 3$ fm for the present work.  However, extending our data out to $b < 6$ fm does not produce a statistically significant change in the double ratio, an insensitivity which is replicated by transport model calculations~\cite{Zhang2014186}.  Considering both statistical and systematic uncertainties, our data is consistent with Ref. ~\cite{famiano06},
  except at the lowest energy data point.  Detailed comparison of the neutron and proton data of the two experiments ~\cite{FamianoPrivate} indicates that the difference lies in the free neutron data, where hardware problems in the previous experiment required large systematic corrections to the neutron spectrum.  The comparison in Figure \ref{fig:ciDR} suggests that the systematic uncertainties of these corrections were probably underestimated in the previous work.   The current data is statistically more precise above 40 MeV than the previous measurement, and extends the range of the measurement from 70 to 100 MeV. The data lie between the SLy4 and SkM* calculations at high energies, but generally agree with the SLy4 calculations better than with the SkM* calculations below 50 MeV. Similarly, the 120 MeV/u data lies between the calculations.  If the trends in these calculations are not modified by other theoretical considerations, this suggests that the magnitude of the effective mass splitting is even smaller than that for the SLy4 mean field and that such a large reduction $m_{p}$ relative to $m_{n}$ as proposed in the SkM* can be ruled out. As shown in Figure 1, however, these calculations do not accurately reproduce all aspects of the data; indicating the need for a thorough evaluation of the theoretical uncertainties in these tentative constraints.

In summary, we have presented new neutron/proton spectral double ratio data from central $^{124}$Sn+$^{124}$Sn and $^{112}$Sn+$^{112}$Sn collisions at two widely separated beam energies.  These measurements provide an increase in accuracy and kinetic energy range compared to previous data at $E_{beam}=50$~MeV/u, and provide new double ratio data at the previously unmeasured beam energy of $120$~MeV/u.  Comparisons to transport theory simulations indicate that these data are precise enough to place constraints on the isovector momentum dependence of the nuclear EoS. The current comparisons suggest that the isovector corrections to the nucleon effective masses are smaller in magnitude than predicted by either the SkM* or SLy4 mean fields. However, additional calculations with other interactions and model assumptions will be required to assess the theoretical uncertainties of these constraints.

We acknowledge the support of the NSCL beam physics and operations staff, Michigan State
University, the Joint  Institute for Nuclear Astrophysics, and
the National Science Foundation via Grants PHY 1102511 and PHY 0822648.

\bibliographystyle{apsrev}
\def\url#1{}

\end{document}